\documentstyle[12pt,graphicx,subfigure]{article}
\begin{document}
\baselineskip=20 pt
\def\l{\lambda}
\def\L{\Lambda}
\def\b{\beta}
\def\a{\alpha}
\def\d{\delta}
\def\g{\gamma}
\def\mphi{m_{\phi}}
\def\dnul{\partial_{\nu}}
\def\dnuu{\partial^{\nu}}
\def\dmul{\partial_{\mu}}
\def\dmuu{\partial^{\mu}}
\def\eps{\epsilon}
\def\hphi{\hat{\phi}}
\def\vphi{\langle \phi \rangle}
\def\mph{m_\phi}
\def\etamunu{\eta^{\mu\nu}}
\def\bfl{\begin{flushleft}}
\def\efl{\end{flushleft}}
\def\bea{\begin{eqnarray}}
\def\eea{\end{eqnarray}}
\begin{center}
{\large\bf   {Higgs pair production due to radion resonance in Randall-Sundrum model:  
prospects at the Large Hadron Collider}}  
 
\end{center}

\vskip 10pT
\begin{center}
{\large\sl {Prasanta Kumar Das}~\footnote{E-mail:pdas @mri.ernet.in} 
}
\vskip  5pT
{\rm
Harish-Chandra Research Institute,\\
Chhatnag Road, Jhusi, Allahabad-211019, India.} \\
\end{center}

\begin{center}
{\large\sl {Biswarup Mukhopadhyaya}~\footnote{E-mail:biswarup@mri.ernet.in} 
}
\vskip  5pT
{\rm
Harish-Chandra Research Institute,\\
Chhatnag Road, Jhusi, Allahabad-211019, India.} \\
\end{center}

\begin{center}
\vspace*{0.25in}
{\bf Abstract}\\
\end{center}
\noindent 
We consider Higgs pair production at the Large Hadron Collider (LHC)
in a Randall-Sundrum scenario containing a radion field. It is shown that
the enhanced effective coupling of the radion to gluons, together with
contributions from a low-lying radion pole, can provide larger event
rates compared to most new physics possibilities considered so far. We
present the results for both an intermediate mass Higgs and a heavy Higgs,
with a detailed discussion of the background elimination 
procedure.

\newpage

\section{Introduction}

The standard electroweak theory still awaits the discovery of the Higgs boson.
After the Large Electron Positron (LEP) collider has set a lower limit of about
114.5 GeV on its mass \cite{TJ}, the responsibility of finding the Higgs now 
rests
mostly on the Lage Hadron Collider (LHC). At the same time, puzzles such
as the naturalness problem make a strong case for physics beyond the standard
model (SM), just around or above the mass scale where the Higgs boson is 
expected to be found. It is therefore of supreme interest to see if the 
collider signals of the Higgs contain some imprint of new physics.
This necessitates detailed quantitaive exploration of a variety of 
phenomenona linked to the production and decays of the Higgs.

In this paper, we have studied pair production of the Higgs boson at the LHC
as a possible channel for uncovering new physics effects.  In particular, we 
show that the rate of such production receives a large enhancement in a class
of theories with extra compact dimensions, namely, the Randall-Sundrum (RS)
type of models containing a radion field.

As has been mentioned above, the large hierarchy between the 
electroweak scale $M_{EW}$ and the Planck scale $M_{\it Pl}$ is  
somewhat puzzling. Whereas theories like supersymmetry and 
technicolour, each with its own phenomenological implications
and constraints,  have addressed this question, theories with extra spatial 
dimensions proposed as a resolution of this problem have recently drawn a 
lot of attention \cite{VR}. Basically, such theories postulate that all 
hitherto known particles and their standard interactions are confined to
our familiar ($3+1$) dimensional spacetime (on a `brane'), while gravity 
propagates in the `bulk' including additional spacelike compact dimensions. 
On compactification,
the ($3+1$) dimensional projection of gravity gives rise to both massless 
and massive graviton modes, where the latter interacts with the standard
particles with a strength that can be perceptible in TeV-scale 
accelerator experiments.

There are two major variants of the above approach. The first one of them
\cite{ADD} proposes a factorizable metric, large compact 
extra dimensions, and a compatification scheme that gives rise to 
a continuum of gravitonic modes on the Brane. The integrated effects
of this continuum in observable processes have been the subject of a large
number of phenomenological studies. However, given the fact that we aim to
have a bulk gravity scale in the $TeV$ range as a natural cut-off to the 
SM (and therefore as a solution to the naturalness problem), 
such an approach fails to explain a fresh hierarchy that it introduces, 
namely, one between the higher dimensional gravity scale and the inverse of 
the large compactification radius. 

The other  approach is the one proposed by Randall and Sundrum \cite{RS}.
In  this model the fifth dimension corresponds to an $S^1/Z_2$ orbifold and 
the (4+1) dimensional world is described by the 
following `warped' metric
\bea
d s^2 = e^{-2 K R_c |\theta|}\eta_{\mu \nu} d x^\mu d x^\nu
- R_c^2 d \theta^2
\eea

\noindent
where $K$ is the bulk curvature constant and $R_c$ is the radius of the extra 
dimension. The theory postulates two $D_3$ branes, one located at $\theta = 0$ 
where gravity peaks and the other at $\theta = \pi$ where the SM
fields reside. The factor $e^{- 2 K R_c |\theta|}$ appearing in the 
metric is known as the warp factor. 

The interesting feature of such a theory is that projections of the
graviton on the brane at ($\theta = \pi$) gives, apart from the usual 
zero mode, a discrete massive tower with spacing on the TeV scale. In 
addition, the coupling of these massive modes to the SM fields is suppressed
not by the Planck mass $M_P$ but by a mass $M_{P} e^{-KR_{c}\pi}$. In fact,
it can be shown that all quantities with the dimension of mass undergo such
an exponential suppression on this brane, thereby offering a tangible 
solution to the naturalness problem with $K R_{c} \simeq 12$, i.e.
with a combination of masses that are not too widely separated.

The length $R_c$ in this scenario is called the brane separation, and 
can be related to the vacuum expectation value (vev) of some modulus 
field $T(x)$ which corresponds to the fluctuations of the metric over the 
background geometry given by $R_c$. Replacing $R_c$ by $T(x)$, we can rewrite 
the RS metric at the orbifold point $\theta = \pi$ as
\bea
d s^2 = g_{\mu \nu}^{vis} d x^\mu d x^\nu - T(x)^2 d \theta^2
\eea

\noindent 
where $g_{\mu \nu}^{vis} = e^{- 2 \pi K T(x)}\eta_{\mu \nu} 
= \left(\frac{\Phi(x)}{f}\right)^2 \eta_{\mu \nu}$. Here
$f^2 = \frac{24 M_5^3}{K}$ and  $M_5$, the $5$-dimensional Planck scale.

One is thus left with a scalar field $\phi(x)$. 
This field is called the the radion field \cite{GRW}. 
Of course, the modulus field has no potential to start with. Thus one needs to 
generate a stable vacuum for $T(x)$ at $R_c$, which in turn can give
$\phi(x)$ a non-zero vev.  This is done  in terms of the Goldberger-Wise
mechanism \cite{GW}, using a bulk scalar field with non-zero vev's
at the two branes, whereupon a potential for the modulus field is generated, and
one ends up with a radion field whose mass ($m_\phi$)
and vev ($\vphi$) are both within the TeV scale. In particular, 
the radion can be lighter than the other low-lying gravitonic degrees
of freedom. Thus it can very well act as the first messenger of
a scenario with compact extra dimensions, and reveal itself in collider 
experiments. Several studies on the observable implications of the
radion are available in the literature 
\cite{Cheung},\cite{DM}.

Here we wish to focus on the pair-production of Higgs bosons mediated by the
radion at hadron colliders. The two features that can be instrumental in 
enhancing the signal in this channel are ($a$) the accessibility of the
radion resonance for $m_{\phi} > 2 m_{H}$, and ($b$) the relatively
enhanced radion coupling with a pair of gluons at LHC energies. Before
we come to the details of the predicted signal, however, let us
briefly recapitulate the various interactions of the radion in a theory of
the above kind.

We list the relevant interactions of the radion with 
the SM fields in the next section. The general features of Higgs 
pair-production via radion are discussed in section 3. Sections 4 and 5
contain discussions on the predicted signals for $m_h < 2 m_W$ and
$m_h > 2 m_W$ respectively. We conclude in section 6.


\section{Effective interaction of radion with the SM fields}

Radion interactions with the SM fields on the TeV brane (i.e. the one 
located at $\theta = \pi$) are governed by $4$-dimensional 
general covariance. The radion essentially couples to the trace of the 
energy-momentum tensor of the SM fields in the following manner:
\bea
{\mathcal{L}}_{\it int} = \frac{\phi}{\vphi} T^\mu_\mu (SM)
\eea
where $\vphi$ is the radion vev. There are phenomenological limits  
on the $m_\phi$-$\vphi$ parameter space, from which  the lower
bound on $\vphi$ can range from
about the electroweak symmetry breaking scale to about a $TeV$, while
$m_\phi$ can in principle be even lighter than $m_W$.

The trace of the energy-momentum tensor of the SM fields is given by
\bea
T^\mu_\mu (SM) = \sum_{\psi} \left[\frac{3 i}{2} \left({\overline{\psi}} 
\g_\mu \dnul \psi - \dnul{\overline{\psi}} \g_\mu \psi \right)\eta^{\mu\nu} 
- 4 m_\psi {\overline{\psi}} \psi\right] - 2 m_W^2 W_\mu^+ W^{-\mu} 
- m_Z^2 Z_\mu Z^\mu \nonumber \\
+ (2 m_h^2 h^2 - \partial_\mu h \partial^\mu h) + ...
\eea

\noindent The photon and the gluons couple to the radion 
via the usual top-loop diagrams; an added source of enhancement
of the coupling is the trace anomaly term \cite{CDJ}. 
This term arises from the fact
the radion couples to the trace of the energy-momentum tensor, which can 
be equated with the four-divergence of the dilatation current associated with
scale invariance of the theory. Scale invariance is preserved at the tree level
in sectors of the theory which are massless and have no dimensionful couplings.
However, quantum corrections can break such invariance, thereby giving
a value to the four-divergence, which is proportional to the relevant 
beta-function.  This term augments the coupling of the radion to the 
trace of he energy-momentum tensor.  In general, the contribution 
of this term to the interaction Lagrangian can be expressed as 
\bea
T^\mu_\mu (SM)^{anom} = \sum_{a} \frac{\b_a(g_a)}{2 g_a} G_{\mu\nu}^a 
G^{a \mu\nu}
\eea
For gluons, $\b_s (g_s)/{2 g_s} = - [\a_s/{8 \pi}]~ b_{QCD}$ where 
$b_{QCD} = 11 - 2 n_f/3$, $n_f$ being the number of quark flavours. 
On the whole, the effective 
$\phi g g$ interaction is given by
\bea
\frac{i \a_s \d_{ab}}{2 \pi \vphi} \left[ b_{QCD} + I_{QCD}  \right] (p_1.p_2 
\eta_{\mu\nu} - p_{1 \nu} p_{2 \mu})
\eea
where $p_1$ and $p_2$ are the $4$-momenta of the gluons $G_\mu^a$ and 
$G_\nu^b$. The function $I_{QCD}$ is dominated by the top quark 
loop, and  is given by
\bea
I_{QCD} = \frac{2}{9} x_t \left[2 + 3 \sqrt{x_t - 1}~ \l(x_t) - 2(x_t - 1)
\l^2(x_t) \right]
\eea
where $x_t=\frac{4 m_t^2}{m_\phi^2}$, and
\bea
\l(x_t) &=& - sin^{-1}\left(\frac{1}{\sqrt{x_t}}\right), 
~~x_t \geq 1 
\nonumber \\
&=& \frac{1}{2}\left[\pi + i~ ln \left(\frac{1 + \sqrt{1 - x_t}}{1 - 
\sqrt{1 - x_t}}\right) \right], ~~x_t < 1
\eea

It is important to note that for $x_t \ge 1$, $n_f = 5$ and hence 
$b_{QCD} = 23/3$, while for $x_t < 1$, $n_f = 6$ and the corresponding
$b_{QCD} = 21/3$.

\section{Higgs pair  production at LHC}

The pair production of a neutral Higgs boson from gluon-gluon fusion at the 
LHC has already been  
studied in the context of the SM \cite{PSZ} as well as several of its 
extensions such  as the minimal supersymmetric standard model (MSSM) 
\cite{BDEMN}. 
As has been already mentioned, there is an expectation of some trace
of physics beyond the SM being found in signatures of the Higgs, and this
can only be possible by studying Higgs production and decays in as
many channels as possible. From this standpoint, the pair production
of Higgs deserves attention, because the predicted rate of such production is
very small in the SM \cite{PSZ}, and thus any excess 
can be interpreted as the signature of
new physics. The mediation of the heavier neutral Higgs in SUSY has been shown
to be the source of some enhancement in a region of the parameter space. 
Theories with extra dimensions, too, have been studied in this context, both
in the ADD and RS models \cite{Kim}, where it has been reported that
the mediation of gravitons can boost the Higgs pair production rates.
What we wish to emphasize, however, is the fact that the presence of a radion 
in the RS context is of particular significance here. This is because 
(a) whereas the graviton resonance in the RS scenario is usually at too
high a mass to be significant, a relatively less massive radion can be
within kinematic reach, and also (b) the enhancement of radion coupling to
a pair of gluons via the trace anomaly term jacks up the contributions.
We have computed the predicted rates for both the cases of
$m_H < 2m_W$  and $m_H > 2m_W$, and analyzed the viability of the
resulting signals with appropriate event selection strategies.
We have used the CTEQ4L parton distribution function, setting the 
renormalisation scale at the radion mass. We have checked that the predicted
results are more or less unchanged if this scale is set, say, at the
partonic subprocess centre-of-mass energy.

\noindent Using Breit-Wigner 
approximation for the resonant production of a radion, one finds the following
expression for the cross-section for $p p \rightarrow \phi 
\rightarrow h h$  (where the dominant subprocess is 
$g g \rightarrow \phi \rightarrow h h$):
\bea
\sigma_{hh}(p p \rightarrow \phi \rightarrow h h) = \int d x_1 \int d x_2~ 
g_1(x_1)~ g_2(x_2) \hat{\sigma}_{hh}(g g \rightarrow \phi \rightarrow h h)
\eea
where,
\bea
\hat{\sigma}_{hh}(g g \rightarrow \phi \rightarrow h h) =
~\frac{\Gamma[\phi \rightarrow gg]}{128~m_\phi^3 \vphi^2}
~\frac{\hat{s} [\hat{s}^2 + 4 m_h^2\hat{s} + 4 m_h^4 ]}
{[(\hat{s} - m_\phi^2)^2 + m_\phi^2 \Gamma_\phi^2]}
\eea
where $\hat{s}$ is centre-of-mass energy for the partonic subprocess. 
$\hat{s} = x_1 x_2 s$, $s(=14~TeV)$ being the proton-proton
centre-of-mass energy, and  $x_1$,$x_2$, the momentum 
fractions of the first and second partons (gluons) respectively.  
$\mph$ and $\Gamma_\phi$ denote the mass and 
total decay width of the radion \cite{DMR}.

Away from the radion resonance, the term proportional to  $\Gamma^2_\phi$
is of little consequence, and the total rate falls as  $1/\vphi^4$ if 
the radion vev is increased.  This is because the decay width of the
radion in any channel is proportional to $1/\vphi^2$. 
Near resonance, on the other hand,
the contribution is dictated by the term with $\Gamma_\phi$ in the
radion propagator. Given the above dependence of the
decay width on  $\vphi$,  the net contribution near resonance
becomes practically independent of the radion vev.

\section{Event selection strategy and results: $m_h <2m_W$}

The signal of Higgs pair production depends on the final states produced by
Higgs decays. We shall broadly consider two mass ranges here, one
corresponding to $m_h< 2 m_W$, and the other, to  $m_h > 2 m_W$.

In the first case, the dominant decay channel for each Higgs is
$H\longrightarrow b\overline{b}$, so that the signal consists of four b-jets.
Normally, such events are beset with a huge background arising from the 
following sources:

\begin{itemize}
\item The QCD production of (${\bar{b}}b)({\bar{b}}b$),

\item The QCD production of  $({\bar{b}}b)(jj)$ with the two non-$b$ jets
misidentified as $b$,

\item The production of two(${\bar{b}}b$) pairs from different partonic
collisions, arising either from the same pair of protons or from
two different pairs. 

\item The electroweak production of $Z({\bar{b}}b)$ and $W({\bar{b}}b)$. 
\end{itemize}

A very relevant discussion of these backgrounds and the cuts that can be used 
to eliminate them can be found in \cite{BDM}. These cuts have been adapted to
our context in this study. Basically, 
the contributions from mistagged b-jets and different partonic collisions
from the same proton pair are relatively small, the `pile-up' effects arising
from different proton pairs can be also handled with the help of effective
b-tagging. For this, the following b-tagging efficiencies
have been assumed in different transverse momentum ($p_T$) ranges\cite{btag}:
\bea
\epsilon_b &=& 0.6, ~~~~~~~~~~~~~~~~~~~~~~~~~~~ for~~~ p_T > 100~ GeV 
\nonumber \\
      &=& 0.1 + p_T/(200 GeV), ~~~~~for~~~40~ GeV \le p_T \le 100~ GeV 
\nonumber \\
      &=& 1.5 p_T/(100 GeV) - 0.3, ~for~~~25~ GeV \le p_T \le 40~ GeV
\eea

\noindent where we also assume that $b$-jets are tagged
only in the pseudorapidity region $\eta_b \le 2$. \\

With such tagging efficiencies, it is found that the 
irreducible backgrounds become manageable with a further set of
event selection criteria, particularly if the Higgs mass is not in 
close vicinity of the Z-mass (a fact that is confirmed by LEP data). The 
exact cuts used in our analysis are listed below.

\begin{itemize}
\item{} {\bf Reconstruction of the Higgs boson mass}: Since we do not 
differentiate between $b$-and $\overline {b}$-jets, there will be 
four $b$ in the final state and hence 
one has three possible ways to pair them up. We choose
those pairs which correspond to the smallest invariant mass difference 
between them. The reconstructed Higgs boson mass is then defined by 
$M_h = [M_{b_1,b_2} + M_{b_3,b_4}]$/2. Keeping in mind the finite resolution
of the reconstruction procedure, this invariant mass  has been smeared
by a Gaussian distribution of
width $\sigma \approx \sqrt{M_h}$. The window for the reconstructed Higgs 
boson mass in which our search is being carried out is
\bea
0.9 m_{h,in} - 1.5 \sigma \le M_h \le 0.9 m_{h,in} + 1.5 \sigma
\eea

\item{} The mass difference between the invariant masses of the two pairs 
are required to be below the width of Gaussian smear: 
\bea
\delta M_h = \left|M_{b_1,b_2} - M_{b_3,b_4} \right| \le 2 \sigma
\eea

\item{} Large azimuthal angles between the two jets of each pair in the 
transverse plane are demanded:
\bea
\delta \phi_{b_1 b_2}, \delta \phi_{b_3 b_4} > 1
\eea

\vspace*{-0.3in}
\newpage
\begin{figure}
\subfigure[]{ 
\label{PictureOneLabel}
\hspace*{-0.7in}
\begin{minipage}[b]{0.5\textwidth}
\centering
\includegraphics[width=\textwidth]{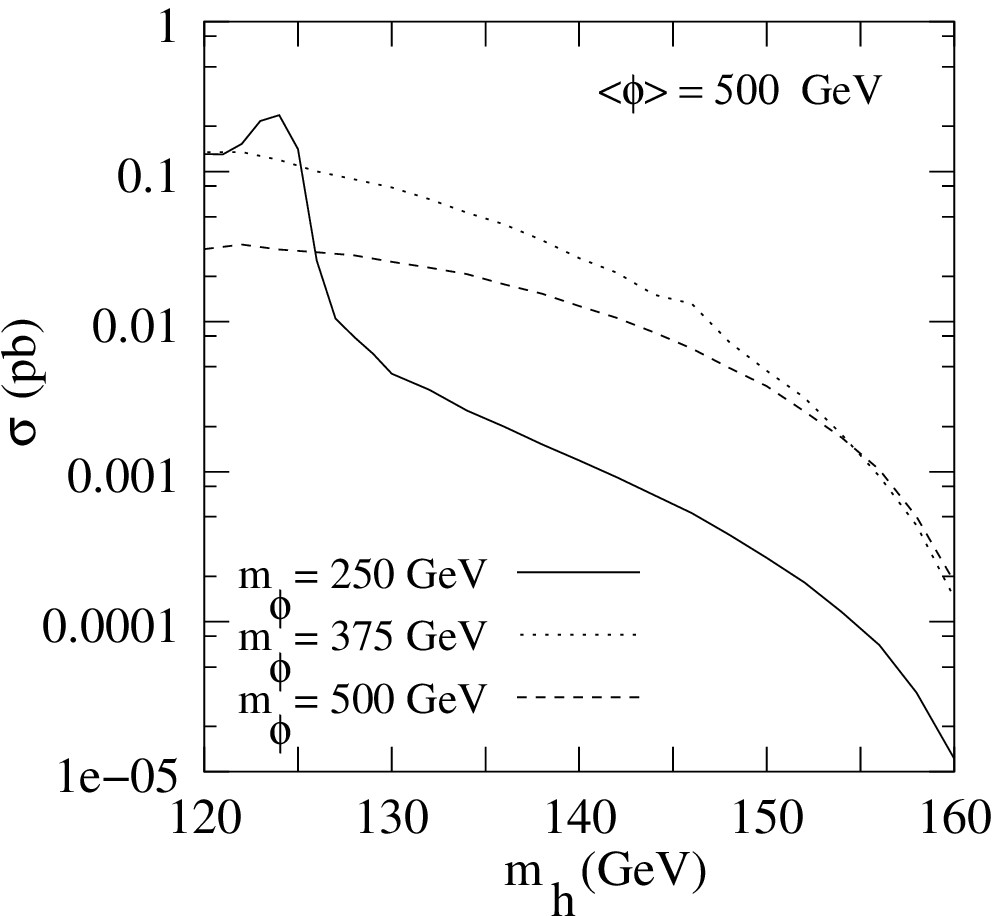}
\end{minipage}}
\subfigure[]{
\label{PictureTwoLabel}
\hspace*{0.3in}
\begin{minipage}[b]{0.5\textwidth}
\centering
\includegraphics[width=\textwidth]{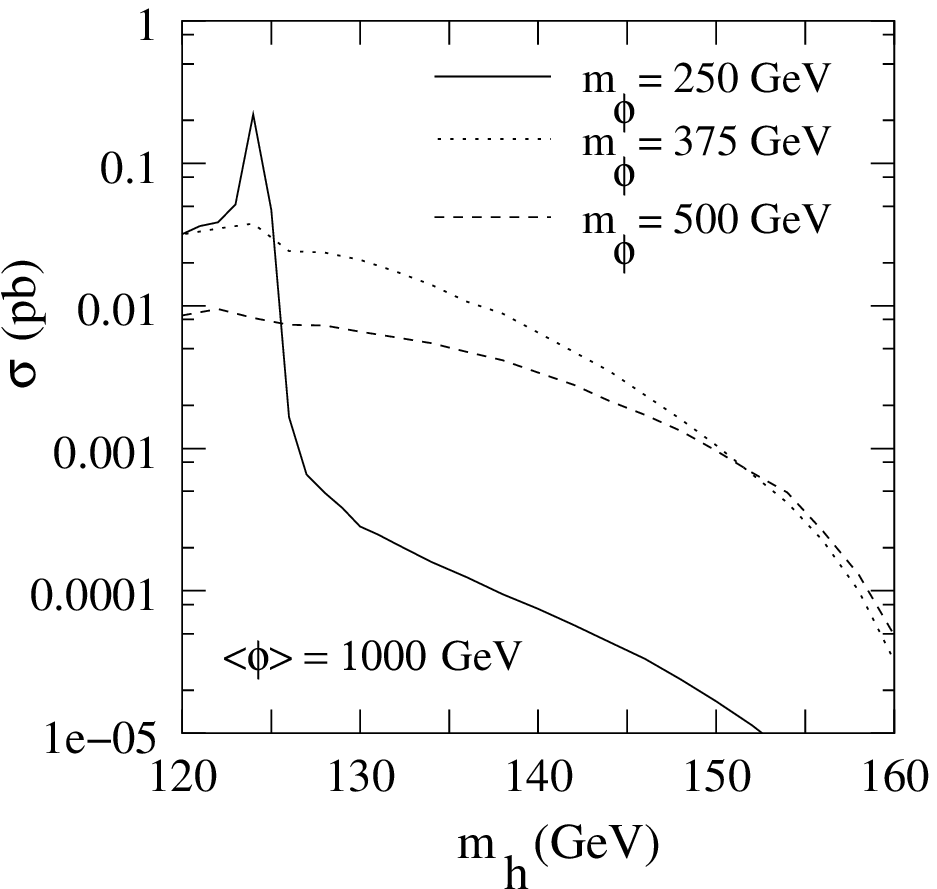}
\end{minipage}}
\caption{\it The Signal cross-section(pb) for the
 process $pp(gg)\rightarrow \phi \rightarrow hh \rightarrow b {\overline{b}}
b {\overline{b}}$ against the Higgs mass for 
$m_\phi$= 250, 375 and 500 GeV and $\vphi$ = 500, 1000 GeV.}
\label{Label_for_both_of_the_figures}
\end{figure}
\item Events where the difference between the angles  
$\delta \phi_{b_1 b_2}, \delta \phi_{b_3 b_4}$ is small are retained:
\bea
\left|\delta \phi_{b_1 b_2} - \delta \phi_{b_3 b_4}\right| < 1
\eea

\item{} The b-jet are subjected to the following $p_T$-cuts: 
\bea
p_{T, min} > M_h/4 ; ~~ p_{T, max} > M_h/4 + 2 \sigma
\eea

\item{} In order to further utilise the hardness  of the b-jets in 
eliminating
backgrounds, a minimum value is imposed on the $4b$ invariant
mass $M_{4b}$:
\bea
M_{4b} > 1.9 M_h - 3 \sigma
\eea
\end{itemize}

\noindent After applying all the above cuts, the cross-section
for the process $pp(gg) \rightarrow \phi \rightarrow h h
\rightarrow b {\overline b} b {\overline b}$ is obtained for
different radion vevs $\vphi$, radion masses $\mph$ and Higgs masses 
$m_h$.
In Figure 1(a,b) we have plotted this cross-section against $m_h$ (GeV)
($\approx m_{h,in}$) for 
$\vphi = 500, 1000$ GeV and for $\mph = 250, 375$ and $ 500$ GeV. 

The first conclusion to draw from the figures is that there  is a 
substantial enhancement of the total rates over what is predicted in the
standard model as well as in the case of MSSM. In addition, it also
exceeds the predicted rates in the ADD-and RS-type models when such models
assume graviton mediation to be the only new effect. This is particularly
evident from the fact that the numbers presented here are {\it after all
the cuts have been applied and the b-tagging efficiency has been folded in}, 
which effectively causes well below one per cent of the signal events to 
survive.

As has been mentioned before, such an
enhancement has
two main sources, namely (a) the availability of the radion resonance,
and (b) the enhanced coupling at the gluon-gluon-radion vertex.   
It can also be noticed by comparing figures 1(a) and 1(b) that the
rates for Higgs masses corresponding to the onset of
resonance are almost independent of the radion vev. Since we are 
showing here only that range of Higgs masses where the decay into
the $b{\overline{b}}$ is dominant, such peaking effect is prominent only
for
\noindent $m_\phi~=~250~GeV$; for the higher values of $m_\phi$, $m_h$  
becomes large enough for the branching fraction in the $b{\overline{b}}$
channel
to drop drastically, before the peak can be reached.

\bfl
{\bf Significance contours}:
\efl

With an integrated luminosity of $100~fb^{-1}$, the above rates indicate
a rather impressive prospect of detecting pair-produced Higgs bosons if
radion mediation is operative. To gauge 
the actual situation, however,
one must also remember that the backgrounds are never totally eliminated,
and thus one must examine how the signal fares compared to
the surviving backgrounds. This is depicted in figures 2(a) and (b)
with contour plots in   $\frac{S}{\sqrt{B}}$,
where, $S$ and $B$ are the number of events corresponding to the signal
and backgrounds respectively with the above luminosity.
In calculating the backgrounds, we have taken both statistical and
systematic effects into account, assuming the systematic uncertainty to be
$2\%$ of the total background and adding it in quadrature to the computed
background itself. The contour plots corroborate our expectation that
the signal really stands out over a large region of

\noindent the parameter space.
Thus the so-called intermediate mass range acquires a high degree of
visibility if a low-lying radion is present in the theory.

\newpage
\begin{figure}  
\subfigure[]{
\label{PictureOneLabel}
\hspace*{-0.7in}
\begin{minipage}[b]{0.5\textwidth}
\centering
\includegraphics[width=\textwidth]{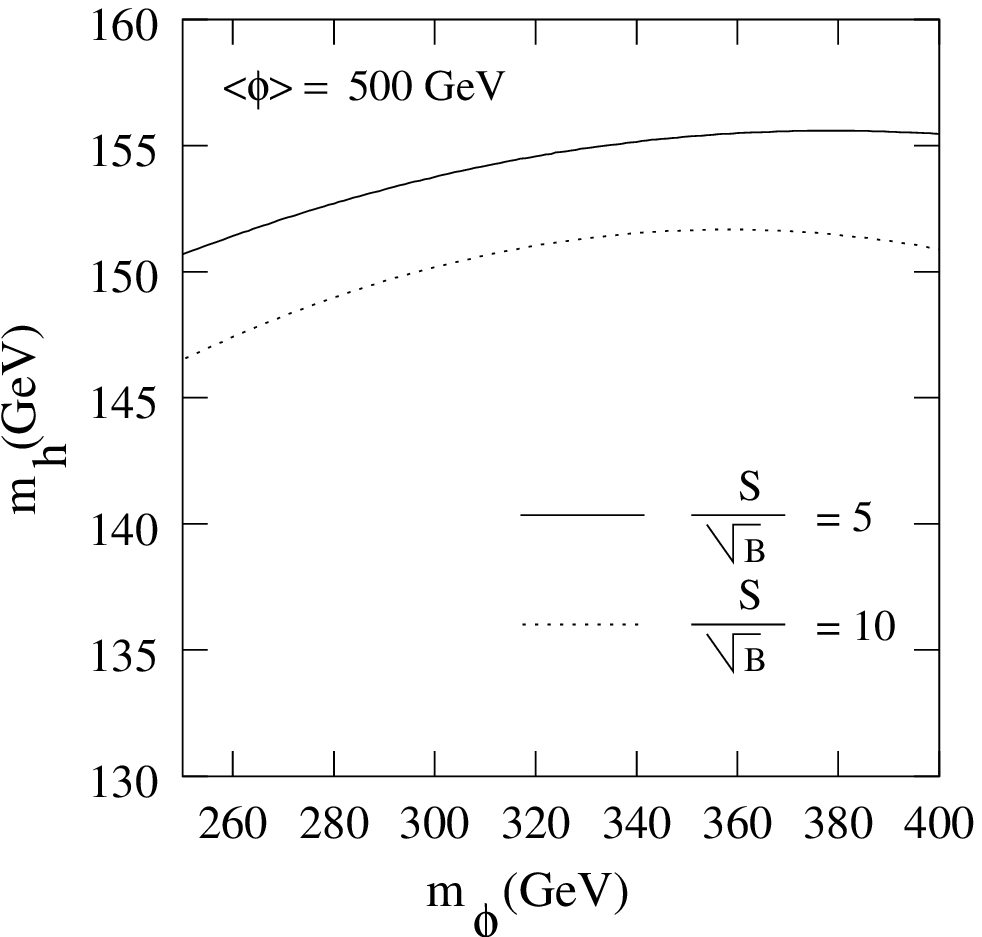}
\end{minipage}}
\subfigure[]{
\label{PictureTwoLabel}
\hspace*{0.3in}
\begin{minipage}[b]{0.5\textwidth}
\centering
\includegraphics[width=\textwidth]{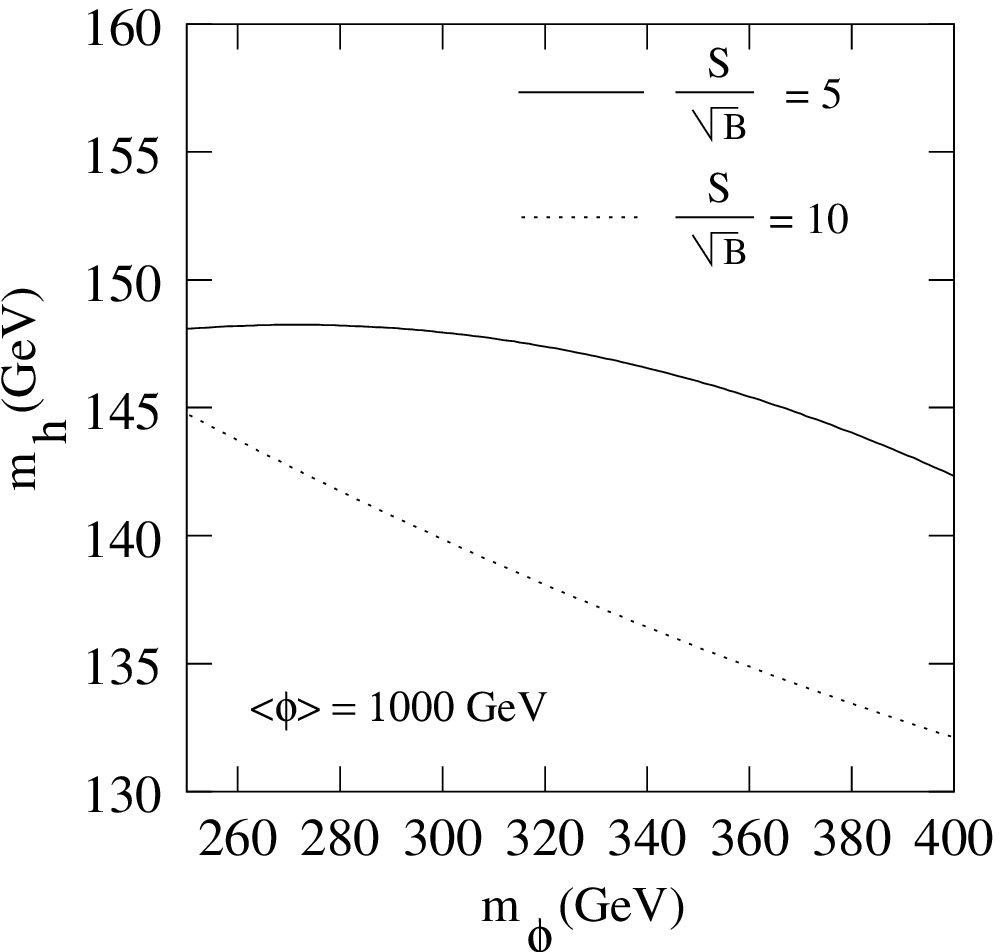}
\end{minipage}}
\caption{\it Contour plots in $m_h$ - $m_\phi$ plane corresponding to
$\vphi$ = 500, 1000 GeV and
$\frac{S}{\sqrt{B}}$ = 5, 10.}
\label{Label_for_both_of_the_figures}
\end{figure}


\section{Event rates for $m_h > 2m_W$}

Let us now consider the case of a somewhat heavier Higgs, which can decay
into a pair of $W$'s or $Z$'s. The signal in such
a case consists in $4W$ or $4Z$ final states, where the SM backgrounds
are negligibly small. In Figures 3 and 4, we have 
plotted the 
cross sections for the processes $pp(gg)\rightarrow \phi \rightarrow hh 
\rightarrow W^+ W^- W^+ W^-$ and $pp(gg)\rightarrow \phi \rightarrow hh 
\rightarrow Z Z Z Z$ (for cases where the  
$W$ and the $Z$ decay into electrons and muons
with total branching ratios of about $0.2$  and $0.06$ respectively)
against the Higgs mass $m_h$ for different $\mph$ and $\vphi$. 
In addition, an average detection efficiency of $75\%$ per lepton has 
been assumed. 
It can be seen from the figures that the $4Z$ final states are very unlikely 
to be seen at the LHC. The $4W$ final state, however, is quite substantial,
especially in view of the absence of backgrounds. One may thus expect
to see events ranging in number from a few tens to several thousands, depending
on the radion mass, so long as the Higgs mass is within approximately $200~GeV$
For a radion mass on the
order of $375~GeV$, a slight peaking behaviour can be seen around

\newpage
\begin{figure}
\subfigure[]{
\label{PictureOneLabel}
\hspace*{-0.7 in}
\begin{minipage}[b]{0.5\textwidth}    
\centering   
\includegraphics[width=\textwidth]{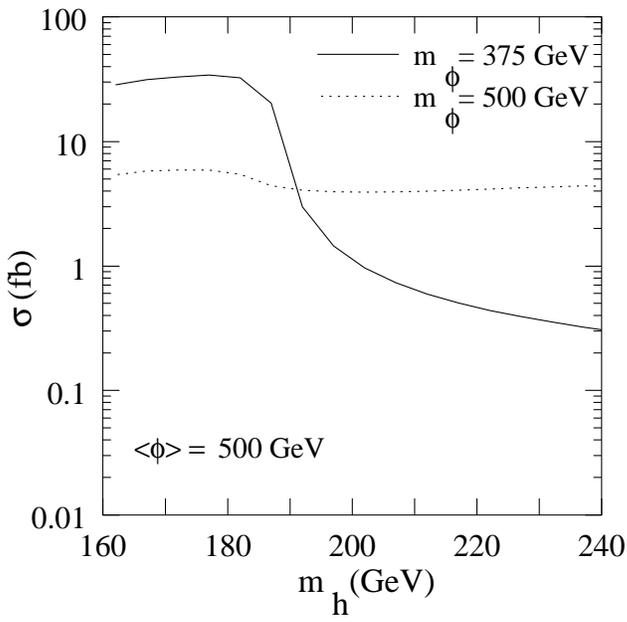}
\end{minipage}}
\subfigure[]{
\label{PictureTwoLabel}
\hspace*{0.3in}
\begin{minipage}[b]{0.5\textwidth}
\centering
\includegraphics[width=\textwidth]{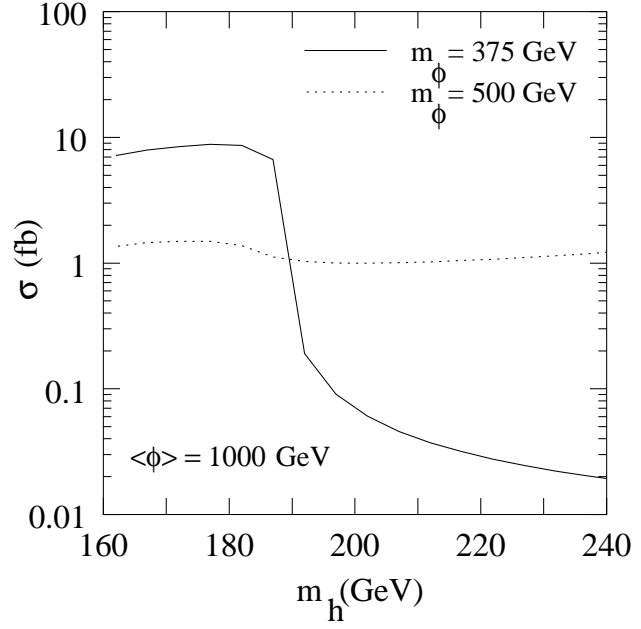}
\end{minipage}}
\caption{\it The Signal cross-section (fb) for the
process $pp(gg)\rightarrow \phi \rightarrow hh \rightarrow W^+ W^-
W^+ W^-$ where each of the produced $W$ decay leptonically
($W \rightarrow l \nu_l, l = e, \mu$) against the Higgs mass
for $m_\phi$= 375 and 500 GeV and $\vphi$ = 500, 1000 GeV.  }
\label{Label_for_both_of_the_figures} 
\end{figure} 
\noindent $m_h\simeq 180~GeV$. For a heavier radion, however, the decay 
width becomes
so large that the peak is washed out.  In this case, the 
$\Gamma_{\phi}^2 m_{\phi}^2$-term in the Breit-Wigner propagator
gives a substantial contribution even where the subprocess centre-of-mass
energy is considerably away from $m_{\phi}^2$. Such a contribution is
responsible for the lack of dependence of the radion vev, following
arguments given earlier.

\section{Summary and conclusion:}

We have looked at the pair production of the Higgs boson
enhanced by  radion mediation at the LHC. For $m_h<2m_W$,
$4b$ final states are investigated. It is found that in spite of
substantial backgrounds to start with, a careful event selection
strategy can lead to a

\newpage
\begin{figure}
\subfigure[]{
\label{PictureOneLabel}
\hspace*{-0.7in}
\begin{minipage}[b]{0.5\textwidth}
\centering
\includegraphics[width=\textwidth]{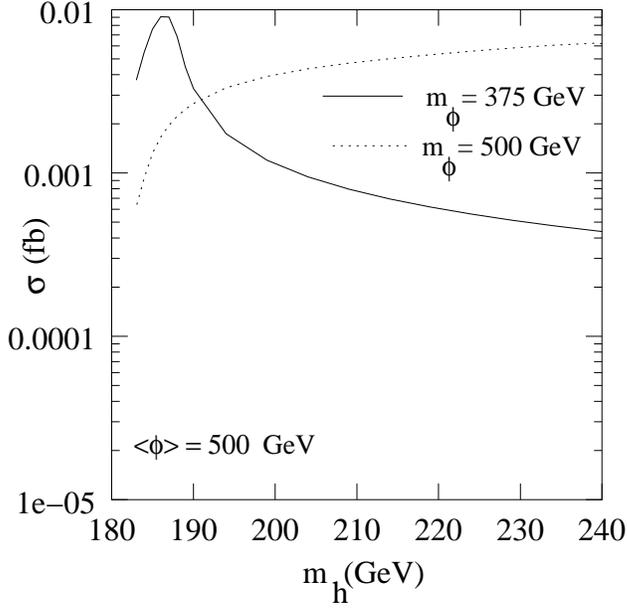}
\end{minipage}}
\subfigure[]{
\label{PictureTwoLabel}
\hspace*{0.3in}
\begin{minipage}[b]{0.5\textwidth}
\centering
\includegraphics[width=\textwidth]{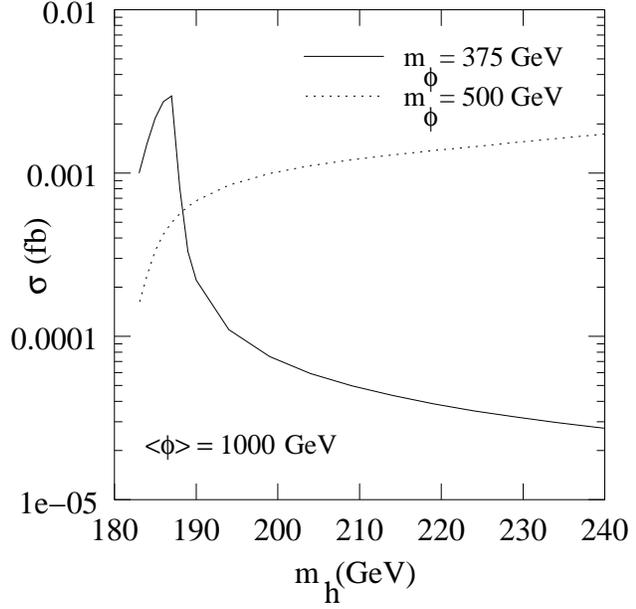}
\end{minipage}}
\caption{\it The Signal cross-section (fb) for the
process $pp(gg)\rightarrow \phi \rightarrow hh \rightarrow Z Z
Z Z$ where each of the produced $Z$ decay leptonically
($Z \rightarrow l^+ l^-, l = e, \mu$) against the Higgs mass for
$m_\phi$= 375 and 500 GeV and  $\vphi$ = 500, 1000 GeV.  }
\label{Label_for_both_of_the_figures}
\end{figure}

\noindent discovery potential at the $5\sigma$ or even
$10\sigma$ level. For $m_h > 2m_W$, on the other hand, one has to depend on 
the $4W$ final states which are almost background-free. There, too,
one should be able to see anything between about 100 and 800 events 
so long as the  radion vev is within a $TeV$ and the Higgs mass 
lies within about $200~GeV$. Thus the presence of a radion can 
boost the very conspicuous 
phenomenon of Higgs pair production, over a large region of the parameter 
space, including the entire range of Higgs mass favoured by precision 
electroweak data. 

\bfl
{\bf Acknowledgment:}
\efl
We thank Uma Mahanta and Sreerup Raychaudhuri
for sharing their radion decay code with us.
PKD would like to thank Partha Konar for
computational help. The work of BM has been partially supported by the
Board of Research in Nuclear Science, Government of India.

\end{document}